\documentclass[prb,twocolumn,showpacs]{revtex4-1}

\usepackage{amssymb}
\usepackage{amsmath}
\usepackage{bm}
\usepackage{hyperref}
\usepackage{graphicx}
\usepackage{ulem}            
\usepackage[svgnames]{xcolor}
\hypersetup{colorlinks=true,linkcolor=blue,citecolor=blue,filecolor=blue,urlcolor=blue}

\newcommand{\Bi}{\bm{i}}
\newcommand{\Bj}{\bm{j}}
\newcommand{\Bell}{\bm{\ell}}

\begin{document}


\title{Electron localization in disordered graphene: multifractal properties of the wavefunctions}
\date{\today}
\author{J.E. Barrios-Vargas}
\author{Gerardo G. Naumis}
\email[E-mail: ]{naumis@fisica.unam.mx}
\affiliation{Depto. de F\'{i}sica-Qu\'{i}mica, Instituto de F\'{i}sica, 
Universidad Nacional Aut\'onoma de M\'exico (UNAM). Apdo. Postal 20-364, 01000, 
M\'exico D.F., M\'exico.}

\begin{abstract}
An analysis of the electron localization properties in doped graphene is performed 
by doing a numerical multifractal analysis. By obtaining the singularity spectrum of 
a tight-binding model, it is found that the electron wave functions present a multifractal 
behavior. Such multifractality is preserved even for second neighbor interaction, which 
needs to be taken into account if a comparison is desired with experimental results. 
States close to the Dirac point have a wider multifractal character than those far 
from this point as the impurity concentration is increased. 
The analysis of the results allows to conclude that in the split-band limit,
where impurities act as vacancies, the system can be well described 
by a chiral orthogonal symmetry class, with a singularity
spectrum transition approaching freezing as disorder increases. 
This also suggests that in doped graphene, localization is in contrast 
with the conventional picture of Anderson localization in two dimensions, 
showing also that the common belief of the absence of quantum percolation in two 
dimensional systems needs to be revised.
\end{abstract}
\keywords{Graphene; Electron mobility}
\pacs{81.05.ue,72.80.Vp,71.23.An,73.22.Pr,72.20.Ee}
\maketitle

Ever since its discovery \cite{Novoselov2004,Novoselov2005}, graphene has been considered as 
an ideal candidate to replace Silicon in electronics \cite{Geim2007}, since this 
first truly two-dimensional crystal has the highest electrical and thermal conductivity known 
\cite{Balandin2008}. However, graphene {\it per se} is not a semiconductor. Several proposals 
have been made to solve this issue\cite{Geim2009}. Experimentally, it has been found that doped graphene 
presents a metal to insulator transition \cite{Bostwick2009} when doped with ${\rm H}$, producing 
a kind of narrow band gap semiconductor. The increase in localization around the Dirac point was 
roughly predicted from an electron wavefunction  frustration analysis in the graphene's underlying 
triangular lattice \cite{Naumis2007,Martinazzo2010,Barrios2011}. 
Such theoretical results were made under the supposition that Hydrogen bonds to the $2p_z$ 
Carbon orbital, and thus impurities act as vacancies \cite{Katoch2010,Peres2010}. This case
corresponds to the split-band limit. This approach has 
been useful to predict localization and the pseudogap size, i.e., the region in which the inverse participation
ratio increases by one order of magnitude \cite{Naumis2007}, in very good 
agreement with experiments\cite{Bostwick2009}, although vacancies and impurities are indeed 
different \cite{LaMagna2009,Deretzis2010}. However, there is a theoretical nuance to the idea 
of having a metal-insulator transition in two dimensions (2D). According to the well known Abrahams's et. al. 
scaling analysis, in 1D and 2D all states are localized for any amount of disorder, excluding 
the possibility of a mobility edge\cite{Abrahams1979}. The experimental and numerical analysis shows that not all 
states are localized, and there exists a kind of mobility edge associated with a pseudogap around the Fermi energy 
\cite{Barrios2011,Barrios2012,Barrios2013}. This means that there is a problem that has not been 
solved. There are two possibilities, either electron-electron interaction produces delocalization 
\cite{Evers2008}, or somehow the Abrahams's et. al. analysis does not completely applies 
to this case. In graphene, electron-electron interaction is very weak \cite{Gonzalez2001}, 
so in principle, the second option is more viable. This possibility has been explored partially 
in a previous publication, since critical states, i.e., states decaying as a power law, were 
observed \cite{Naumis2007,Barrios2012}. The possibility of having such states
has been around in old studies concerning the possibility of having quantum percolation in 2D 
\cite{Meir1989,Soukoulis1991}, 
or in symmetry breaking analysis of graphene \cite{Evers2008}. In this case, the extra symmetries 
of the graphene's lattice makes the problem different from a  generic 2D Fermi gas 
\cite{Barrios2012}. 
Such idea has been found by making a random matrix analysis of 
broken symmetries in the Dirac Hamiltonian when disorder is introduced \cite{Evers2008}. In this analysis,
a classification of the symmetries leads to different classes of universalities in the transition.
Although all these points have been around for a while, few numerical results are available analyzing
such questions \cite{Markos2012}. Here we provide such analysis, proving that multifractal states are present in doped graphene, which in the split band limit turns out to be in a chiral orthogonal symmetry broken class 
approaching a freezing singularity spectrum.

Let us consider doped graphene as a honeycomb lattice with substitutional impurities placed at random 
with a uniform distribution. The corresponding $\pi$  orbital one electron tight-binding Hamiltonian is \cite{Wallace1947},
\begin{align}
\mathcal{H}=-t\sum_{\langle \Bi,\Bj\rangle} |\Bi\rangle \langle\Bj|-t'\sum_{\langle\langle \Bi,\Bj\rangle\rangle} 
|\Bi\rangle \langle\Bj|+
\varepsilon \sum_{\Bell} |\Bell\rangle \langle \Bell|\,. 
\end{align}

The first sum is over nearest neighbors, with $t=2.79\,{\rm eV}$ the hopping energy
\cite{Reich2002}. The second sum is carried over second neighbors. Here we will consider two cases, $t'=0\,{\rm eV}$ which
is the most studied Hamiltonian, and $t'=0.68\,{\rm eV}$, that gives a much better approximation to real graphene\cite{Reich2002}. 
The idea is to study the effects of including second neighbors interaction in the problem of localization.
The third sum is over impurity sites with self-energy $\varepsilon$. 
The number of impurities sites, $N_{\rm imp}$, is determined by the concentration 
$C=N_{\rm imp}/N_{\rm T}$, where $N_{\rm T}$ is the total sites on the honeycomb lattice.

Around the Fermi energy, this model presents exponentially localized wavefunctions appear, as has been documented in a previous publication by our group\cite{Barrios2012}. Such
wavefunctions are in agreement with the Abrahams's et. al. theorem. Here, we will concentrate on wavefunctions that are above the pseudomobility edge, which has been  proven to have a size given by \cite{Naumis2007,Barrios2012}   $\Delta \approx t \sqrt{6C}$. 

\begin{figure}[h] 
\begin{tabular}{c}
   \includegraphics[width=\linewidth]{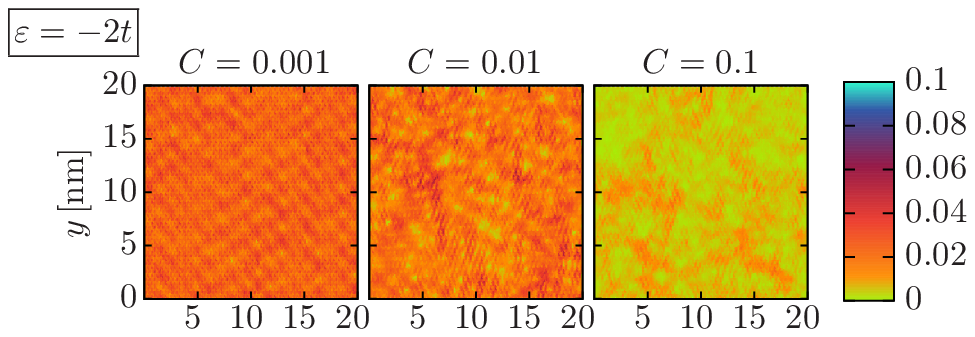} \\
   \includegraphics[width=\linewidth]{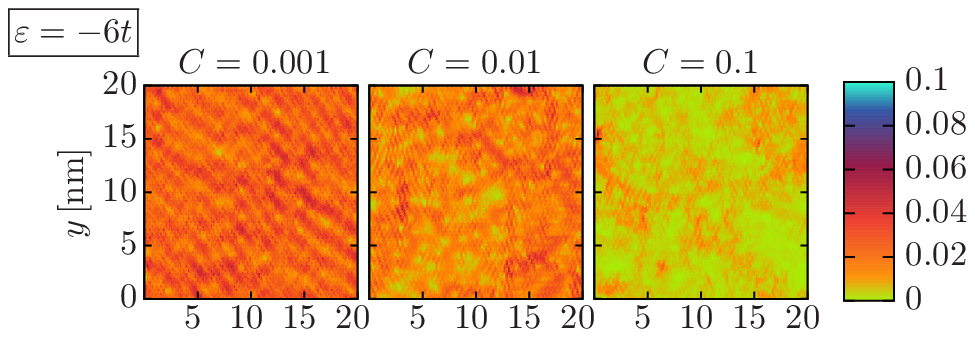} \\
   \includegraphics[width=\linewidth]{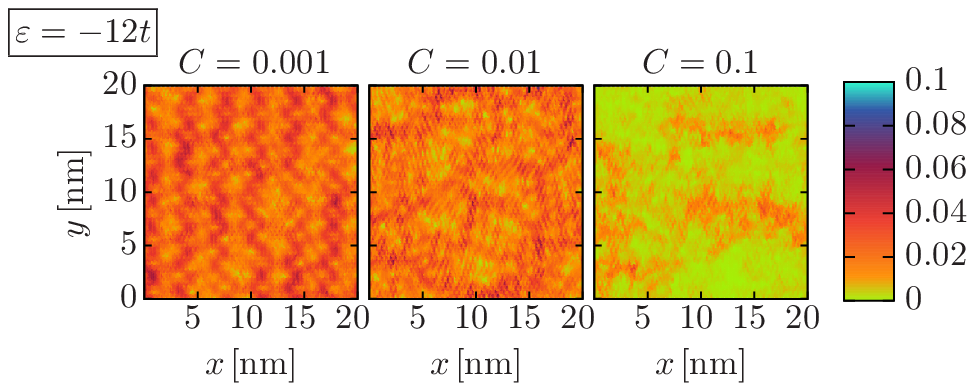} 
\end{tabular}
   \caption{Electron wavefunction amplitude close to the Dirac point, but outside the region where the 
participation ratio begins to decrease, corresponding to $E=E_{\rm D}-0.6t$, for different concentrations of disorder 
($C=0.001$, $C=0.01$ and $C=0.1$) and different impurity self-energy ($\varepsilon=-2t$, $\varepsilon=-6t$ and 
$\varepsilon=-12t$), using a lattice of $N_{\rm T}=18432$ sites.} 
   \label{Wave1}
\end{figure}

In Figure \ref{Wave1}  we present a sample of the observed wavefunctions $\psi(\bm{r})$ which solve the Schr\"odinger equation $H\psi(\bm{r})=E\psi(\bm{r})$ for an energy close to the Dirac  energy, $E_{\rm D}$, but outside the region where the participation ratio begins to decrease \cite{Naumis2007}. As can be seen, there is a progressive change of the localization as $C$ or $\varepsilon$ increases. However,  although disorder increases up to a concentration of $C=0.1$, no evident localization center appears. This suggest to perform a multifractal analysis to confirm this hypothesis.

The multifractal analysis can be performed as follows\cite{Thiem2013}. The system with area $A=N \times N$, where $N$ 
is a number of primitive cells per side\footnote{Then, the number of total sites in the sample is $N_{\rm T}=2N^2$.},
is divided into $B=N^2/L^2$ boxes of linear size $L$. On a given box $b$, the probability of finding the electron is given by,
\begin{align}
\mu_b(\psi,L)=\sum_{\bm{r} \,\in {\rm box}\, b} \vert \psi(\bm{r})\vert ^2\,.
\end{align}

A measure is built by normalizing the moments of this probability,
\begin{align}\label{DistMom}
\mu_b(q,\psi,L)=\frac{\big[\mu_b(\psi,L)\big]^q}{P(q,\psi,L)}\,,
\end{align}
where $P(q,\psi,L)$ is,
\begin{align}
P(q,\psi,L)=\sum_{b=1}^{B}\big[\mu_b(q,\psi,L)\big]^q \,.
\end{align}

The mass exponent of the wave function can be obtained using,
\begin{align}
\tau_q(\psi)=\lim_{\delta\rightarrow 0} \frac{P(q,\psi,L)}{\ln \delta} \,,
\end{align}
where $\delta$ is the ratio $N/L$. The fractal dimensions $D_q$ is introduced via\cite{Evers2008} $D_q=\tau_q/(q-1)$ . In an insulator $D_q=0$ while for a metal $D_q=d$. In multifractal cases, $D_q$ is a function
of $q$. 

Using the previous equation \eqref{DistMom}, it is possible to find the singularity spectrum $f(\alpha_q)$, which is basically the fractal dimension of the set of points where the wavefunction behaves as 
$\vert \psi(\bm{r})\vert^{2} \sim  L^{-\alpha_q}$.\cite{Halsey1986} For a finite system, the number of such points scales as $L^{f(\alpha_q)}$. This singularity spectrum $f(\alpha_q)$ is obtained by observing that, 
\begin{align}
\alpha_q(\psi)=\lim_{\delta \rightarrow 0} \frac{A(q,\psi,L)}{\ln \delta} \,,
\end{align}
where $A(q,\psi,L)$ is,
\begin{align}
A(q,\psi,L)=\bigg\langle \sum_{b=1}^{B}\mu_b(q,\psi,L) \ln \mu_b(1,\psi,L) \bigg\rangle\,.
\end{align}

\begin{figure} 
\centering
   \includegraphics[width=\linewidth]{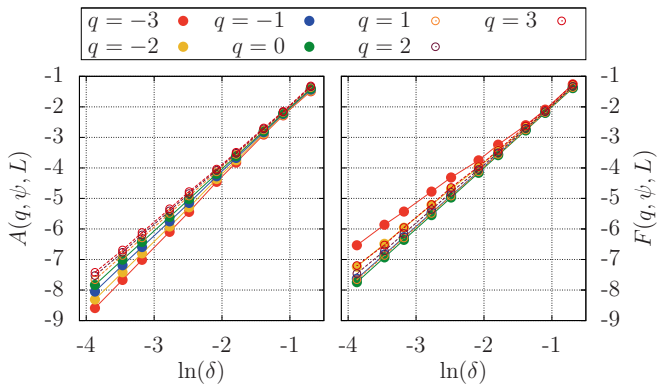} 
   \caption{(Color online) Left panel: Typical behavior of $A(q,\psi,L)$.
Right panel: Typical behavior of $F(q,\psi,L)$. Both graph parameters are $E=E_{\rm D}-0.2t$, $\varepsilon=-6t$ and $C=0.0001$.} 
   \label{AFgraf}
\end{figure}

Here $\langle ... \rangle$ denotes the arithmetic average over many realizations of disorder. In Figure~\ref{AFgraf},
the typical behavior of $A(q,\psi,L)$ is plotted as a function of $\delta$ for several $q$. For each $q$, a straight line can be fitted in order to get the slope and from there obtain $\alpha_q$.
The singularity spectrum $f(\alpha_q)$ is obtained as,
\begin{align}
f(\alpha_q)=\lim_{\delta \rightarrow 0} \frac{F(q,\psi,L)}{\ln \delta}\,,
\end{align}
where $F(q,\psi,L)$ is a kind of entropy information,
\begin{align}
F(q,\psi,L)=\bigg\langle \sum_{b=1}^{B} \mu_b(q,\psi,L) \ln \mu_b(q,\psi,L) \bigg\rangle\,,
\end{align}
where $F(q,\psi,L)$ can be calculated as done with $A(q,\psi,L)$, 
since all points fall in a straight line as a function of $\delta$ (Figure~\ref{AFgraf}).

\begin{figure}[h] 
\centering
   \includegraphics[width=\linewidth]{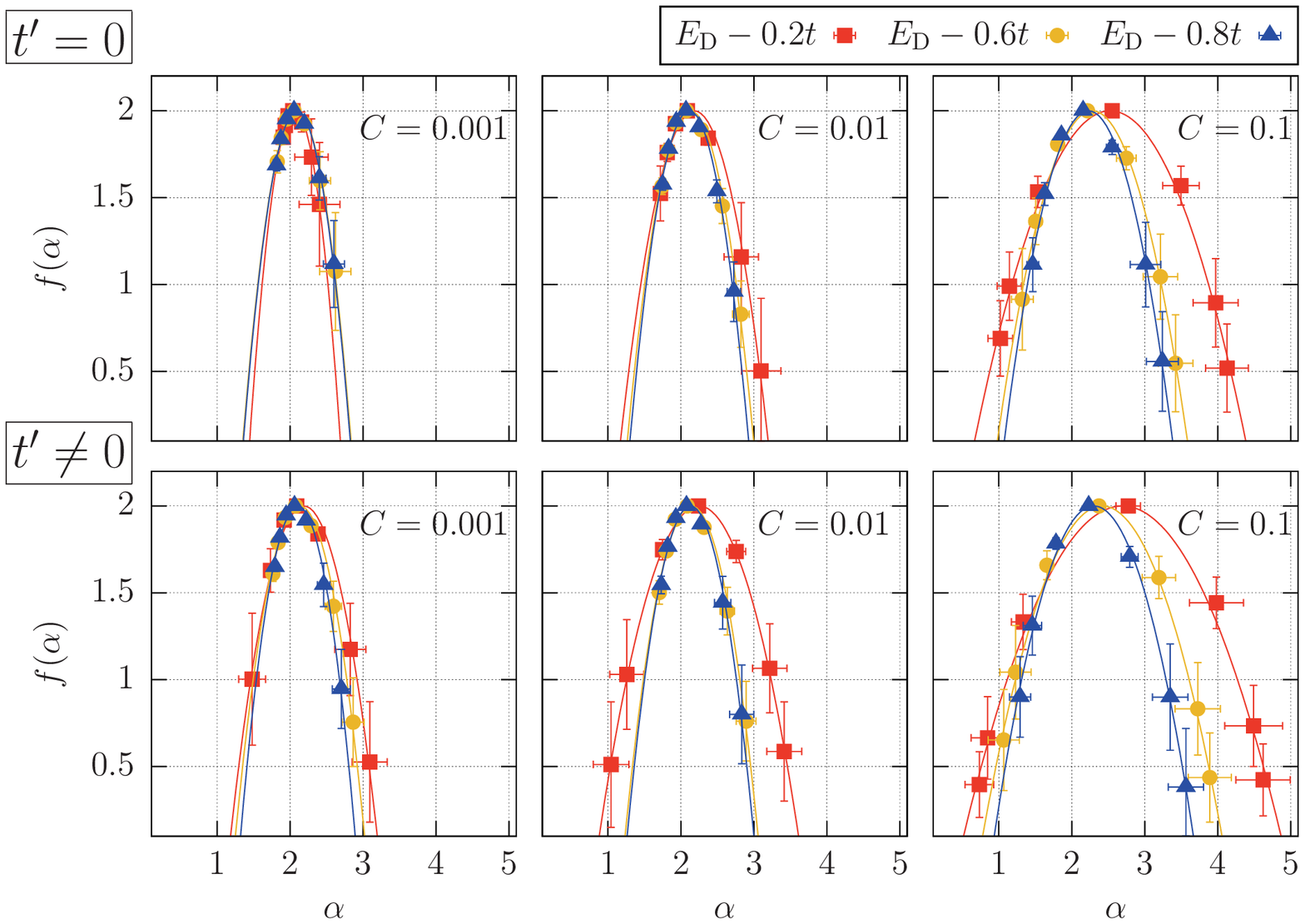} 
   \caption{(Color online) Singularity spectrum for three representative states. A state near the Dirac energy $E_{\rm D}-0.2t$ (red 
   squares), $E_{\rm D}-0.4t$ (gold circles), and far from $E_{\rm D}$ at $E_{\rm D}-0.8t$ (blue triangles). Here
   $\varepsilon=-6t$. The first row corresponds to the nearest-neighbor interaction $t'=0$, while the second corresponds to
   the next-nearest-neighbor interaction $t'\neq 0$. The solid lines are obtained by fitting the data with a parabola. 
Notice how the state far from the Dirac point 
    always has a more pure fractal behavior than their counterparts. The plot was made using an average over $30$ lattices of $N_{\rm T}=18432$ sites.} 
   \label{falfapv}
\end{figure}

\begin{figure}[h] 
\centering
   \includegraphics[width=\linewidth]{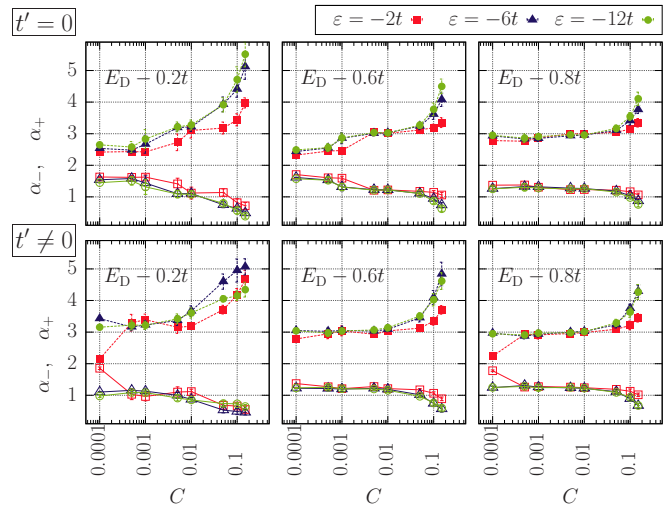} 
   \caption{(Color online) Evolution of the roots of $f(\alpha)$ ($\alpha_{\rm -}$ open symbols and $\alpha_{\rm +}$ 
filled symbols) as a function of the impurity concentration for the states at $E=E_{\rm D}-0.2t,\,E_{\rm D}-0.6t,\,E_{\rm D}-0.8t$, 
using different values for $\varepsilon$. Top panel corresponds to the nearest-neighbor interaction $t'=0$ and bottom panel 
corresponds to the next-nearest-neighbor interaction $t'\neq 0$.} 
   \label{fspread}
\end{figure}

As a control of the involved analysis, we have verified that for pure graphene $\varepsilon=0$, the singularity spectrum converges to the point $f(2)=2$, which is exactly the expected value for Bloch states in $2$ dimensions.
In Figure~\ref{falfapv} we present the singularity spectrum for three states at 
$E=E_{\rm D}-0.2t,E_{\rm D}-0.6t,E_{\rm D}-0.8t$, chosen
to represent states near and far from the Dirac energy. This plot was made considering $\varepsilon=-6t$ for an average of $30$ lattices of $N_{\rm T}=18432$ sites. First of all, 
a convex parabola is observed, showing a typical weak multifractal behavior. This proves that multifractal states are present, which was the main hypotesis of this work. We have 
verified that this multifractal
behavior is observed for many other states, as well as for different set of disorder parameters $C$ and $\varepsilon$ (see below). 
Also, the figure shows the tendency for states near the Dirac point to have a wider multifractal distribution, while states far 
from $E_{\rm D}$ tend to have a more pronounced mono-fractal character, as expected from a frustration analysis of the underlying triangular symmetry of the lattice \cite{Barrios2011}. In fact, it is interesting to compare with the 
2D-limit of a strongly 
disordered system, in which exponentially localized states are observed, with a spectrum that converges to the 
points $f(0)=0$ and $f(\infty)=2$. This tendency for states near the Dirac point 
was also confirmed in the present work as disorder was increased, since the parabola $f(\alpha)$ tends 
to reach the origin, and at the same time spreads over bigger values of $\alpha$, indicating a tendency for localization. 
Such spreading can be quantified by looking at the roots of $f(\alpha)$, i.e. $f(\alpha_{-})=f(\alpha_{\rm +})=0$ with $\alpha_{\rm -}<\alpha_{\rm +}$; as shown in Figure~\ref{fspread}. In this figure, one can observe how for small disorder, the roots tends to collapse, as expected for states which are closer to a Bloch behavior. 
The spreading becomes bigger for high $C$ and $\varepsilon$. Another important feature is the  value $\alpha_0$ for which $f(\alpha_0)=2$, corresponding to the maximal value of $f(\alpha)$. Due to this property, in the limit $N\to\infty$, for almost all  points the amplitudes scale as $|\psi(\bm{r})|^2 \sim L^{-\alpha_0}$, 
where $\alpha_0>2$. This confirms that eigenfunctions 
follow a power law decay, a point that was already discussed in great detail in a previous publication\cite{Barrios2012}

Since in real graphene the next-nearest-neighbor interaction is important, in Figure~\ref{falfapv} we present $f(\alpha)$ for the same three states and parameters
 including $t'$. Notice how states close to the Dirac point are more localized, while the state far from the Dirac point is nearly equal to its counterpart in Figure~\ref{falfapv}. This 
more pronounced behavior can be roughly understood as an increase in the effective dimensionality  of the problem due to the 
next-nearest-neighbor interaction. Also, in Figure~\ref{fspread}, we present the spreading of the $f(\alpha)=0$ when next-nearest-neighbor 
interaction is included.

\begin{figure} 
\centering
   \includegraphics[width=\linewidth]{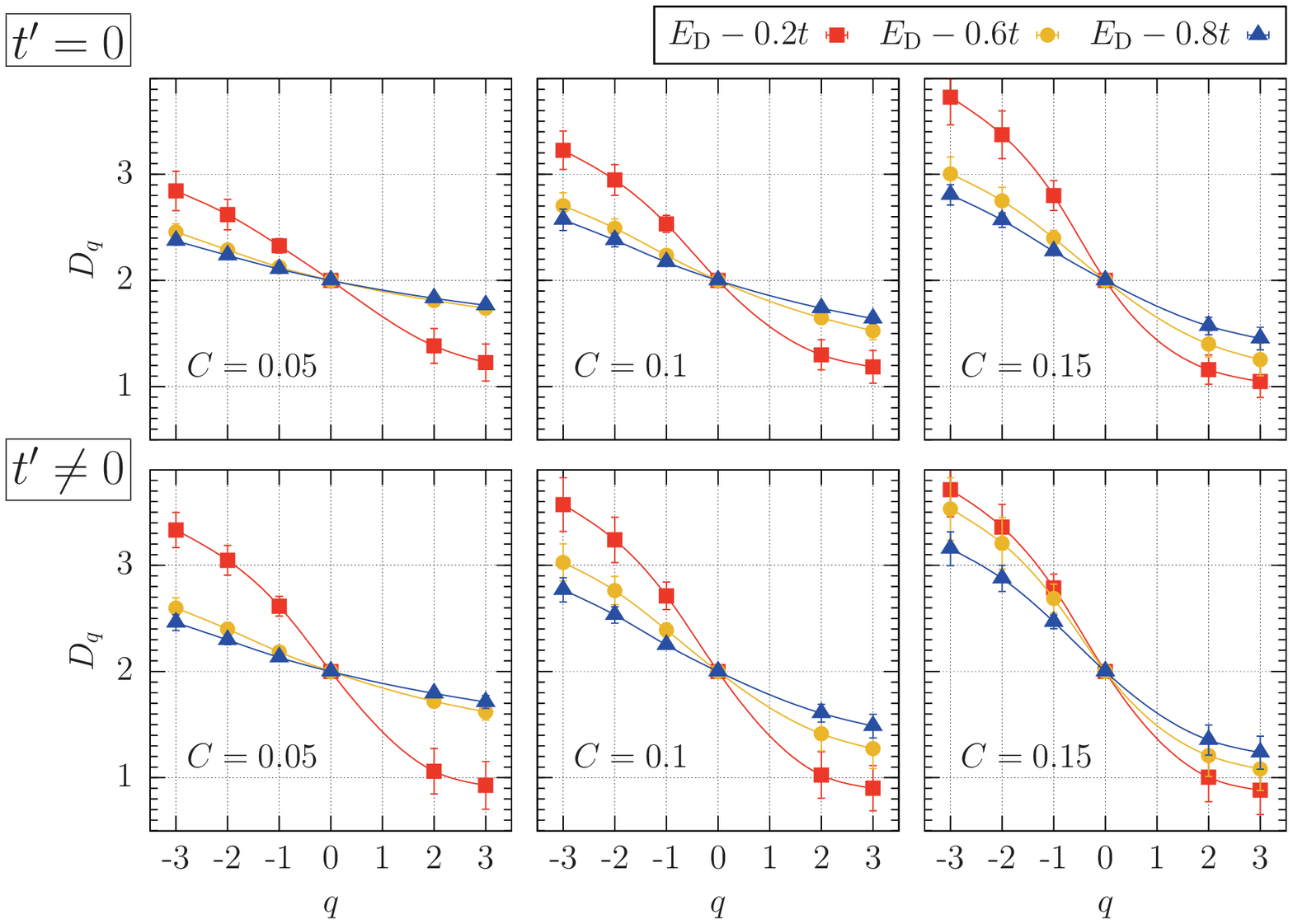} 
   \caption{(Color online) $D_q$ as a function of $q$ for three representative states using $\varepsilon=-6t$, obtained by averaging over $30$ lattices of $N_{\rm T}=18432$ sites.} 
   \label{DqFigura}
\end{figure}

\begin{figure} 
\centering
   \includegraphics[width=\linewidth]{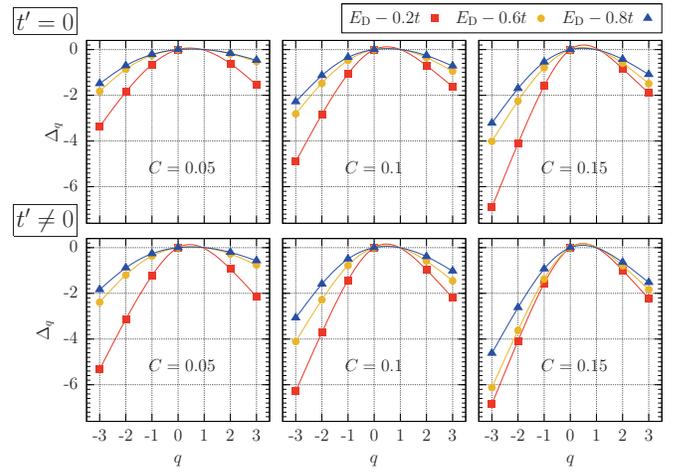} 
   \caption{(Color online) $\Delta_q$ for three different states using the same set of parameters of the previous plot. Notice the asymmetry with respect to $q=0$} 
   \label{Deltaq}
\end{figure}

In Figure~\ref{DqFigura}, we present the evolution of $D_q$ for the same three states using $\varepsilon=-6t$ at different $C$. 
The less dispersion of $D_q$ for $E=E_{\rm D}-0.8t$ indicates again a less dispersed mono-fractal character of states far from 
$E_{\rm D}$. An interesting quantity to look for, is the anomalous dimensions $\Delta_q$ defined as,
\begin{align}
\Delta_q\equiv \tau_q-2(q-1)\,,
\end{align}
which separates the normal part in such a way that distinguish the metallic phase from the critical point and determines the scale dependence of the wave function correlations. In Figure~\ref{Deltaq} we present the corresponding result for $\Delta_q$ using the same set of parameters used in previous equations. The most important feature to remark in the plot is the absence of symmetry around $q=0$. As we will see below, this allows to classify the type of the broken symmetries when disorder is included. It is worthwhile mentioning that the value $\Delta_2$ gives the  decaying exponent of the wave function amplitude correlations\cite{Abrahams2010}, i.e.,  
\begin{align}
N^{4}\langle  |\psi^{2}({\bf r})\psi^{2}({\bf r'})| \rangle \sim \bigg(\frac{|{\bf r}-{\bf r'}|}{N}\bigg)^{-\eta}\,,
\end{align}
with $\Delta_2=-\eta$. This confirms again that wave functions close to the Dirac point decay faster, since from 
Figure~\ref{Deltaq}, $|\Delta_2|$ is bigger than the corresponding values for states far from 
$E_{\rm D}$.   

\begin{figure} 
\centering
   \includegraphics[width=\linewidth]{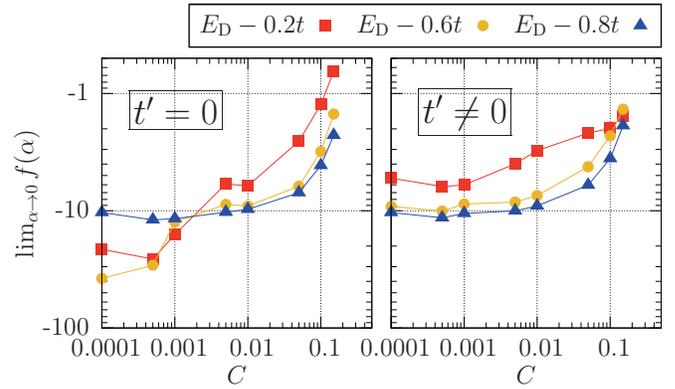} 
   \caption{(Color online) Values of $f(\alpha)$ at $\alpha=0$ for different states as a function of $C$ for $\varepsilon=-12t$.} 
   \label{fatzero}
\end{figure}

One can extract more information on the nature of the singularity spectrum when disorder is included by calculating the behavior of $f(\alpha\to 0)$ as a function of the impurity concentration. Three kinds of behaviors are known\cite{Evers2008}, a) no singularity when $\lim_{\alpha \rightarrow 0}f(\alpha)=-\infty$, b)
termination when $\lim_{\alpha \rightarrow 0}f(\alpha)={\rm constant}$ and c) freezing when $\lim_{\alpha \rightarrow 0}f(\alpha)=0$. In Figure~\ref{fatzero} we present the results obtained from our simulation, where the $f(0)$ was obtained by looking at the intersection of the fitted curves of $f(\alpha)$. From Figure~\ref{fatzero}, the 
results indicates that all plots corresponds to case b), with a tendency to reach freezing as the disorder increases.

In the split-band limit, the previous results seem to confirm the random matrix ensemble symmetry analysis in graphene's context \cite{Ostrovsky2007}.  According to this approach, multifractality arises due to a breaking of the Dirac Hamiltonian symmetries. Such Hamiltonian contains a bipartite (chiral) symmetry which is inherited from the fact that two atoms are present in the 
graphene's unitary cell. Each kind of perturbation leads to different broken symmetries. In the case of vacancies, which is akin to our case $\varepsilon\gg t$, the chiral, temporal and isospin symmetries are preserved. The resulting Hamiltonian belongs to the chiral ortogonal symmetry (BDI class), resulting in a Gade-Wigner type of theory, characterized by lines of fixed points for the renormalization group and non-universal conductivity\cite{Evers2008}. In this class, the localization transition corresponds to the case of termination for weak disorder and freezing for strong disorder, a feature that is confirmed by our plots of $\Delta_q$ 
(Figure~\ref{Deltaq}) and $f(\alpha=0)$ (Figure~\ref{fatzero}). Furthermore, the absence of symmetry in $\Delta_q$ with respect to $q$ in Figure~\ref{Deltaq} means that the results do not belong to the Wigner-Dyson class, which confirms the fact that the system preserves the chiral symmetry. 

Notice that for finite $\varepsilon$, the disorder introduced by us does not correspond to the BDI class, since
chirality is not preserved, basically because there are elements in the diagonal of the Hamiltonian matrix which leads to a  non symmetric spectrum around the Dirac point. However, in the limit $\varepsilon\gg t$, the symmetry is
reestablished in the Carbon band \cite{Naumis2007,Barrios2011-e}, leading to a chiral symmetry Hamiltonian. In other words, impurities with $\varepsilon\gg t$ can be also treated by setting all bonds connected to an impurity with a hopping parameter $t=0$. Thus, our model reduces to the analysis made by Ostrovsky et al. \cite{Ostrovsky2007} in the case $\varepsilon\gg t$, where the orthogonal symmetry class implies strong localization.

In conclusion, we have given numerical evidence of the multifractal nature of wave functions in doped graphene, which was predicted before using symmetry analysis in the split band-limit \cite{Ostrovsky2007,Evers2008}. Such conclusion is important to understand the scaling of the conductance as a function of the system size. This leads to many open questions, as for example, the scaling behavior of doped graphene's nanoribbons.

This work was supported by DGAPA-UNAM project under Grant IN-102513. Computations were done at supercomputer NES of DGTIC-UNAM.
\bibliographystyle{apsrev4-1}
\bibliography{biblioMultifractal}
\end{document}